\documentclass[aps,prb,reprint,groupedaddress]{revtex4-2}
\usepackage[dvipsnames]{xcolor}
\usepackage{graphicx}
\usepackage{amsmath,amssymb}
\usepackage{bm}
\usepackage{braket}
\usepackage{lipsum}
\usepackage{makerobust}
\usepackage{hyperref}
\usepackage{comment}

\newcommand*\circled[1]{\tikz[baseline=(char.base)]{
    \node[shape=circle,draw,inner sep=1pt] (char) {#1};}}
\MakeRobustCommand\circled

\newcommand{\makeauthor}[2]{\newcommand{#1}[1]{{%
  \sffamily\color{#2}{%
    \bfseries\begingroup\escapechar=-1\edef\x{\endgroup\string#1}\x:%
  } ##1}}%
  \MakeRobustCommand#1}
\makeauthor{\eric}{Plum}
\makeauthor{\themba}{ForestGreen} 
\makeauthor{\sr}{magenta}
\makeauthor{\cole}{Dandelion}
\makeauthor{\Fig}{red}

\begin{document}

\renewcommand{\vec}[1]{\bm{#1}}
\newcommand{\up}{{\uparrow}}
\newcommand{\dw}{{\downarrow}}
\newcommand{\pa}{{\partial}}
\newcommand{\pd}{{\phantom{\dagger}}}
\newcommand{\bs}[1]{\boldsymbol{#1}}
\newcommand{\add}[1]{{{\color{black}#1}}}
\newcommand{\todo}[1]{{\textbf{\color{red}ToDo: #1}}}
\newcommand{\tbr}[1]{{\textbf{\color{red}\underline{ToBeRemoved:} #1}}}
\newcommand{\eps}{{\varepsilon}}
\newcommand{\nn}{\nonumber}
\def\ie{\emph{i.e.},\ }
\def\eg{\emph{e.g.},\ }
\def\ea{\emph{et. al.}\ }
\def\cf{\emph{cf.}\ }

\newcommand{\brap}[1]{{\bra{#1}_{\rm phys}}}
\newcommand{\bral}[1]{{\bra{#1}_{\rm log}}}
\newcommand{\ketp}[1]{{\ket{#1}_{\rm phys}}}
\newcommand{\ketl}[1]{{\ket{#1}_{\rm log}}}
\newcommand{\braketp}[1]{{\braket{#1}_{\rm phys}}}
\newcommand{\braketl}[1]{{\braket{#1}_{\rm log}}}

\graphicspath{{./}{./figures/}}



\title{Effect of impurities and disorder on the braiding dynamics of Majorana zero modes}

\author{Cole Peeters}
\affiliation{School of Physics, University of Melbourne, Parkville, VIC 3010, Australia}
\author{Themba Hodge}
\affiliation{School of Physics, University of Melbourne, Parkville, VIC 3010, Australia}
\author{Eric Mascot}
\affiliation{School of Physics, University of Melbourne, Parkville, VIC 3010, Australia}
\author{Stephan Rachel}
\affiliation{School of Physics, University of Melbourne, Parkville, VIC 3010, Australia}

\date{\today}

\begin{abstract}
Impurities and random disorder are known to affect topological superconducting phases and their Majorana zero modes (MZMs). In particular, it is a common assumption that disorder negatively influences the braiding dynamics of MZMs. Recently, it was shown, however, that random disorder can also stabilize or even increase topological phases. Here, we investigate quantitatively how a single impurity can lead to braiding errors. We show that the impurity increases, in most scenarios, the dynamical hybridization of the MZMs, reducing the braiding performance. In addition, we show how random disorder, i.e., impurities on all lattice sites but with different strengths, affects braiding. As for the static case, we observe a window of opportunity where random disorder decreases the average energy of a braid, and thus improves braiding outcomes. This window of opportunity is, however, limited due to an increase of diabatic effects in the presence of disorder. Nevertheless and contrary to physical intuition, disorder can in certain situations be beneficial and improve braiding outcomes.
\end{abstract}

\maketitle

%
%

\section{Introduction}

Majorana zero modes (MZMs) have emerged as key components in the quest for topological quantum computing \cite{nayak2008,alicea2012,dassarma2015}.
These exotic quasiparticles offer a unique advantage due to their topological protection.
As long as MZMs remain sufficiently separated in space, they are protected by a topological energy gap, which shields them from local perturbations.
One of the remarkable features of MZMs is their ability to store quantum information non-locally using pairs of MZMs.
Exploiting the non-Abelian statistics of MZMs, quantum operations can then be performed on this encoded information through a process known as braiding, where the worldlines of MZMs are braided in space-time.

However, disorder has long been recognized as a significant challenge in the practical realization of MZMs.
Any real-world system is bound to contain some form of disorder, whether from impurities\,\cite{woods2021}, imperfections in the interface\,\cite{takei2013}, or other environmental factors\,\cite{vuik2016,goto2024}.
Theoretical studies have extensively explored the impact of disorder on MZMs \cite{akhmerov2011,brouwerPRB2011,brouwerPRL2011,dassarma2012,liu2012,pikulin2012,degottardiPRL2013,degottardiPRB2013,sau2013,adagideli2014,hegde2016,gergs2016,awoga2017,pekerten2017,lieu2018,crawford2020,pan2020,ahn2021,dassarma2021,woods2021,boross2022,dassarma2023,dassarma2023a,goto2024,pan2024}, showing that these modes are remarkably robust to disorder, even when the disorder strength exceeds the superconducting gap by a significant margin.
A particularly interesting avenue is the stabilization and growth of the topological phase in extensions of the Kitaev chain due to disorder\,\cite{gergs2016,lieu2018}.
All these studies, however, primarily focus on static systems, where the disorder and MZMs remain in fixed positions. A key question remains: what happens when MZMs move through a disordered potential, as they must in any braiding operation? The possibility that such motion could introduce errors in the braiding process has yet to be fully understood. In this article, we address this issue by investigating how disorder affects the non-Abelian properties of MZMs during braiding. Specifically, we analyze two cases: the interaction of MZMs with individual impurities and their movement through randomly disordered potentials.
Another aspect of braiding dynamics are diabatic effects: when MZMs are moved too quickly they tend to excite into higher energy states, similar to Landau-Zener transitions, which leads to drastic braiding errors.

Quite generally there are two limits on the braiding speed, sometimes referred to as the {\it speed limit} of braiding \cite{karzig2013}. (i) There is a lower bound due to diabatic effects, set by the inverse gap size $\Delta_{\rm eff}$.
(ii) Hybridization effects can increase the energies of the MZMs away from zero energy, $E_{\rm hyb}$, which causes the fidelity of the braid to oscillate with frequency $\sim \hbar/E_{\rm hyb}$, setting the upper bound.
The speed limit \add{for the braid time $T$} can thus be expressed as\,\cite{brouwerPRL2011,karzig2013,sanno-21prb054504,hodge2024}
\begin{equation}
\label{speed-limit}
\hbar/\Delta_{\rm eff} \ll T \ll \hbar/E_{\rm hyb}\ .
\end{equation}
As we will see below, impurities and disorder can affect both the lower and upper bounds.

%
%
\section{Methods}

We perform braids on a T-junction, composed of three Kitaev chains \cite{kitaev2001} \add{labelled $l=1,2,3$}.
The time-dependent Hamiltonian for each chain is given by
\begin{align}
    H_l
    = -&\sum_{i=1}^{L} \mu_{l,i}(t) c_{l,i}^\dagger c_{l,i}^{\phantom{\dagger}} \nonumber\\
    -&\sum_{i=1}^{L-1} \left(
        \tilde{t} c_{l,i}^\dagger c_{l,i+1}^{\phantom{\dagger}}
        + \Delta_l c_{l,i}^\dagger c_{l,i+1}^\dagger
        + {\rm H.c.}
    \right)
\end{align}
with hopping $\tilde{t}$ as well as pairing amplitudes $\Delta_l$ and chemical potential $\mu_{l,i}$ on chain $l$ (i.e., leg $l$ of the T-junction).
The chains are connected at the junction site,
\begin{align}
    H_0 &= -\mu_0(t) c_0^\dagger c_0^{\phantom{\dagger}}
    - \sum_{l=1}^3 \left(
        \tilde{t} c_{0}^\dagger c_{l,1}^{\phantom{\dagger}}
        + \Delta_l c_{0}^\dagger c_{l,1}^\dagger
        + {\rm H.c.}
    \right).
\end{align}
The entire time-dependence is contained in $\mu(t)$ where $t$ is time. 
An impurity on chain $l$ at site $i$ is introduced by adding a constant term $Vc_{l,i}^\dagger c_{l,i}^{\phantom{\dagger}}$ to the Hamiltonian where $V$ is the strength of the onsite potential.  
Disorder is introduced by placing a random impurity at each site, the strength of which is sampled from a uniform distribution on the interval $[-\frac{w}{2},\frac{w}{2})$.

\begin{figure}[t!]
    \centering
    \includegraphics[width=\columnwidth]{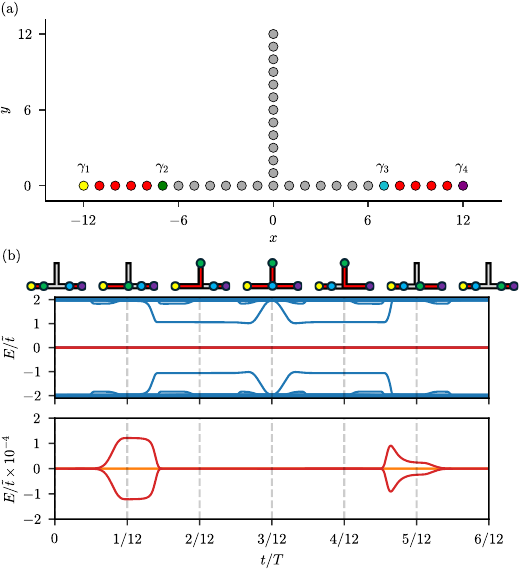}
    \caption{(a) Initial configuration of topological (red) and trivial (grey) segments of a 37-site T-junction (each leg with $L=12$ sites). The four MZMs $\gamma_i$ 
    (colored) are at the ends of the topologial regions. (b) \add{Representative} energy levels of the bulk states (blue) and MZMs (red) throughout the braid \add{in the absence of disorder}, with diagrams showing the state of the T-junction at seven equal spaced times. Bottom: Zoom into the energies of the MZMs.}
    \label{fig:braid}
\end{figure}

We consider throughout the paper the braid of a $\sqrt{\rm X}$-gate or X-gate operating on a single logical qubit; it consists of four MZMs $\gamma_1, \ldots, \gamma_4$ located at the ends of two topological chain-segments placed on a T-junction, as shown in Fig.\,\ref{fig:braid}\,(a).
Exchanging $\gamma_2$ and $\gamma_3$ once corresponds to a $\sqrt{\rm X}$-gate, and exchanging them twice to an X-gate.
Braiding is done by sequentially tuning the chemical potential $\mu_i$ at each site, following the protocol described in Refs.\,\cite{alicea2011,sekania2017,mascot2023}, as illustrated in Fig.\,\ref{fig:braid}\,(b).
The braid is split into three steps.
In ``step 1'' from time $t=0$ to $t/T=1/6$, $\gamma_2$ moves from the left chain to the top chain; in ``step 2'' from $t/T=1/6$ to $t/T=1/3$, $\gamma_3$ moves from the right chain to the original location of $\gamma_2$; and in ``step 3'' from $t/T=1/3$ to $t/T=1/2$, $\gamma_2$ moves to the original location of $\gamma_3$.

We define the qubit states in terms of the occupation of the two Majorana states. There are two sub-spaces that the system could be in, defined by whether the number of fermions is odd or even. This parity is a conserved quantity as the system is superconducting, therefore there is no interaction between the two sub-spaces and they can be considered independently. For systems with an even parity we define the logical states as $|0\rangle_L= |00\rangle_{\rm phys}$ and $|1\rangle_L= |11\rangle_{\rm phys}$, while for the odd parity $|0\rangle_L= |01\rangle_{\rm phys}$ and $|1\rangle_L= |10\rangle_{\rm phys}$. \add{The physical states $\ket{n_1 n_2}_{\rm phys}$ correspond to the parities of the two topological segments.}

The fidelity of the braid describes the overlap between a pair of logical states $a$ and $b$ after $b$ is time evolved through the braid, and is defined as $\mathcal{F}_{ab}=|\langle a(t=0)|b(t=T)\rangle|^2$. As an X-gate should transform $|0\rangle\rightarrow|1\rangle$ and vice versa, a good braid's fidelity should be $\mathcal{F}_{01}=\mathcal{F}_{10}=1$ and $\mathcal{F}_{00}=\mathcal{F}_{11}=0$.
We simulate braids using the method in Ref.\,\onlinecite{mascot2023}. We estimate qubit and braiding errors as outlined in Ref.\,\onlinecite{hodge2024}.

%
%

\section{\label{sec: single impurities}Single Impurities}

We will consider adding a single static impurity to the system, but let us first discuss the clean case.
The instantaneous energies during the braiding protocol are shown in Fig.\,\ref{fig:braid}(b).
We have chosen the chemical potential of the topological regions, $\mu_{\rm{topo}}$, to be $0.05\tilde{t}$ and the trivial chemical potential, $\mu_{\rm triv}$, to be $10\tilde{t}$. 
This, combined with $|\Delta|=\tilde{t}$, means that that the Majorana wavefunctions are more localized within the topological region compared to the trivial region, leading to the predominant energy splitting coming from hybridization between $\gamma_2$ and $\gamma_3$ when they are brought close together. 
That is, the hybridization occurs through the trivial region\,\cite{hodge2024}.
This effect is approximately $10^4$ times larger than the splitting arising from the interaction between $\gamma_1$ and $\gamma_2$, as well as $\gamma_3$ and $\gamma_4$. 
We stress that using longer chains with a smaller gap will have similar results.

In step 1, as $\gamma_2$ moves onto the top chain, a subgap state appears near $E\approx \tilde{t}$.
This subgap state appears since the left and top chain are both in the topological phase but the $\Delta_l$'s have a phase difference\,\cite{alicea2011}.
Since this state is sufficiently far from zero energy, there should be no transitions between this state and the MZMs, as long as the braid is adiabatic.
Zooming in near zero energy [Fig.\,\ref{fig:braid}\,(b)], there is a small energy splitting at $t=T/12$ and $5T/12$ of the braiding process since $\gamma_2$ and $\gamma_3$ are closest during these steps.
The shape of the energy splitting is different for steps 1 and 3 due to an asymmetry in the ramping procedure of the chemical potential.
This small but finite energy splitting leads to oscillations in the braid outcome\,\cite{cheng2011,hodge2024} and sets an upper bound on the total braid time.
Therefore, to reduce braiding errors, it is best to minimize the average energy, $\bar{E}$\,\cite{hodge2024}, which is defined to be the integral of the instantaneous energies $E_n(t)$ of the MZMs over the full braid time $T$, $\bar{E}=\frac{1}{T}\int_0^T [E_1(t)+E_2(t)] dt$.
\begin{figure}[t!]
    \centering
    \includegraphics[width=\columnwidth]{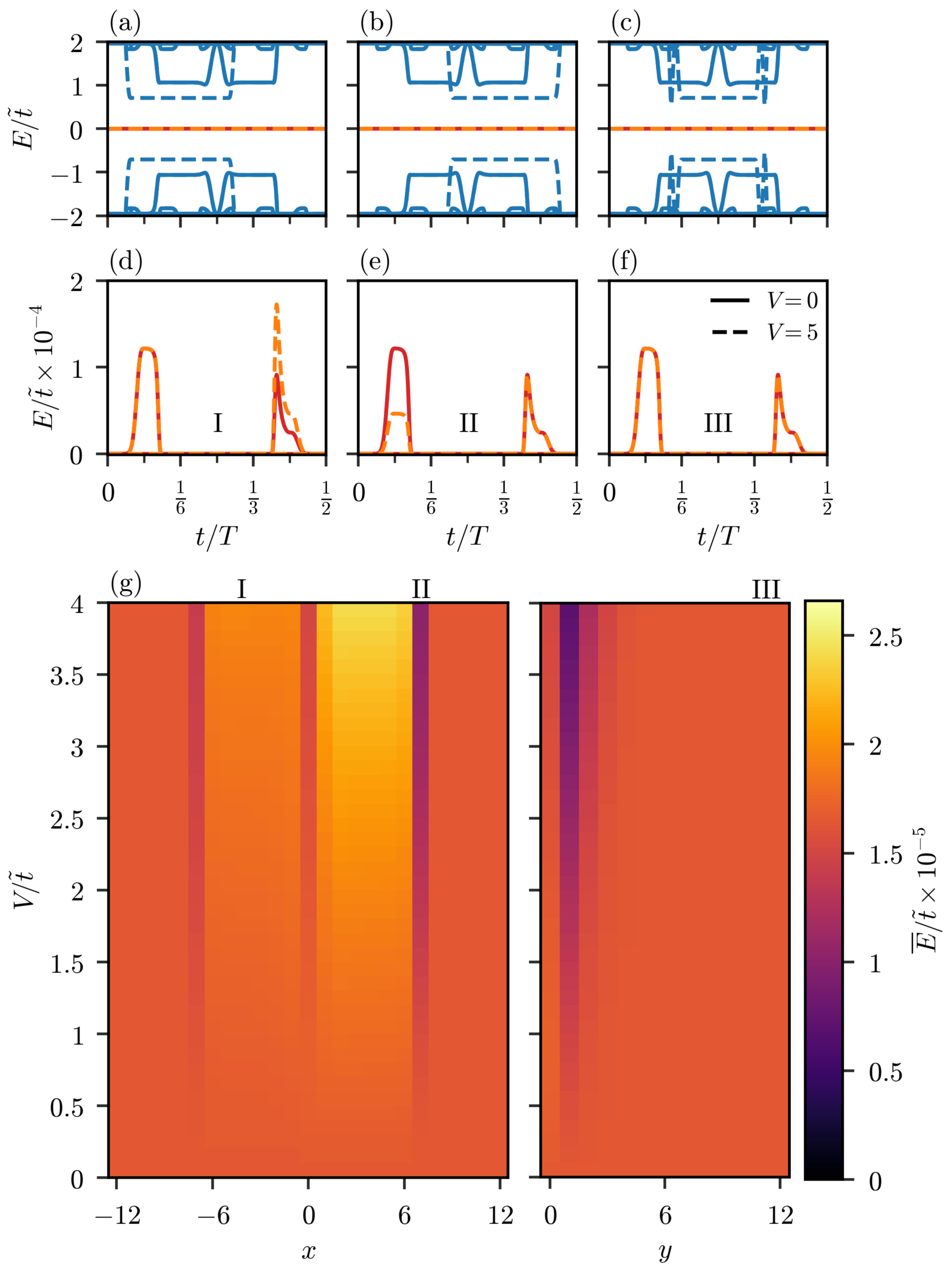}
    \caption{(a--c) Instantaneous energies for half the X-gate (i.e., $\sqrt{\rm X}$-gate) for $V=0$ ($V/\tilde{t}=5$) shown as solid (dashed) lines.
    The impurity is located in (a) at $\vec{r}_{\rm I}=(-4,0)$, in (b) at $\vec{r}_{\rm II}=(7,0)$ and in (c) at $\vec{r}_{\rm III}=(0,11)$.
    (d--f) Zooming in close to zero energy of panels (a--c). Note that the solid lines in (a-f) are identical, and only the dashed lines differ, describing the three different impurities.
    (g) Average energy $\bar{E}$ for a single impurity on every site of the T-junction as a function of disorder strength $V/\tilde{t}$.}
    \label{fig:YSR-E}
\end{figure}

When an impurity is added, it will induce a Yu-Shiba-Rusinov (YSR) bound state inside the gap\,\cite{balatsky-06rmp373,sau2013,kaladzhyan2019} if the chain is in the topological phase.
In the trivial phase, a low-energy bound state appears when $V \approx \mu_{\rm triv}$ (see App.\,\ref{app:subgap} for details). In the topological phase, this YSR state can come arbitrarily close to zero energy as $V \rightarrow \infty$, and it is possible to have accidental transitions to the YSR state during braiding for sufficiently large impurity potential, even in the limit $\hbar/\Delta_{\rm eff} \ll T \ll \hbar/E_{\rm hyb}$.
We will first consider the case where the YSR state's energy is sufficiently large.

When the impurity is placed on the top chain, a YSR state appears for step 2 (dashed line in Fig.\,\ref{fig:YSR-E}(\add{c})).
The impurity does not affect the hybridization between MZMs (Fig.\,\ref{fig:YSR-E}\,(f)) since the impurity is \add{in the topological region when it is located between $\gamma_2$ and $\gamma_3$}.
If the impurity is instead placed on the left chain between the initial locations of $\gamma_2$ and $\gamma_3$, the YSR state appears during step 1 (dashed line in Fig.\,\ref{fig:YSR-E}(\add{a})). 
In Fig.\,\ref{fig:YSR-E}\add{(d)}, we see the energy splitting increase during step 3 but unchanged during step 1.
This happens because the impurity is in the topological region before the MZMs start to hybridize in step 1.
However, in step 3, the impurity is in between $\gamma_2$ and $\gamma_3$ when they are hybridizing most strongly.
The impurity enhances the hybridization, increasing the energy splitting (step 3 in Fig.\,\ref{fig:YSR-E}(\add{d})).

This can be explained by considering the dependence of the Majorana localization length $\xi$ on the chemical potential. As the energy splitting is proportional to $e^{-r/\xi}$\,\cite{kitaev2001} increasing $\xi$ leads to the observed larger energies. As is well known for the Kitaev chain, $\xi(|\mu|\to 2\tilde{t})\to\infty$ which is the critical point where the MZMs are uniformly distributed over the chain. As the chemical potential moves further away from this critical point, $\xi$ decreases in both topological and trivial regions. Suppose that $\xi=\xi_i$ is a local property of the chain. 
If an impurity $V_i$ adjusts the chemical potential of 
that site such that $\mu_i \to \tilde{\mu}_i = \mu_i-V_i$ gets closer towards $|\mu|= 2\tilde{t}$ \add{(note that $\mu_i>2\tilde{t}$ in the trivial region)}, the coherence length will increase, leading to larger energies. This could be explicated by using transfer matrices.

An interesting case occurs when the impurity is located at the initial site of $\gamma_3$.
The impurity effectively shortens the topological segment in the right chain.
This increases the distance between $\gamma_2$ and $\gamma_3$ in step 1, which causes the energy splitting to decrease in Fig.\,\ref{fig:YSR-E}\,(e). A similar situation occurs when the impurity is on the initial site of $\gamma_2$. \add{The resulting decrease in $\bar{E}$ is clearly visible in Fig.\,\ref{fig:YSR-E}\,(g) at $x=\pm 7$.}

The effect of an impurity located at each site on the average energy, $\bar{E}$, is shown in Fig.\,\ref{fig:YSR-E}\,(g). 
In general, we see that an impurity on the top chain or in the initial topological segments does not affect the average energy.
This is because it does not affect the distance between $\gamma_2$ and $\gamma_3$, the main source of hybridization in this braid. An impurity located between the initial locations of $\gamma_2$ and $\gamma_3$ generally enhances the hybridization and increases the average energy.
The enhancement affects the right chain more strongly due to the asymmetry of the braiding protocol (which can be seen in the larger energy splitting in step 1 compared to step 3 of Fig.\,\ref{fig:braid}\,(b)).
Placing an impurity at the junction site has a unique effect.
In this case, as a MZM is transported through the junction, it will avoid this site (the wave function is suppressed at the site), effectively increasing the distance between MZMs and decreasing the hybridization, as observable in the center (i.e., $x=0$) of Fig.\,\ref{fig:YSR-E}\,(g). We observe similar reductions of $\bar{E}$ e.g.\ on site $y=1$; the detailed analysis is left for future investigation.

\begin{figure}[t!]
    \centering
    \includegraphics[width=\columnwidth]{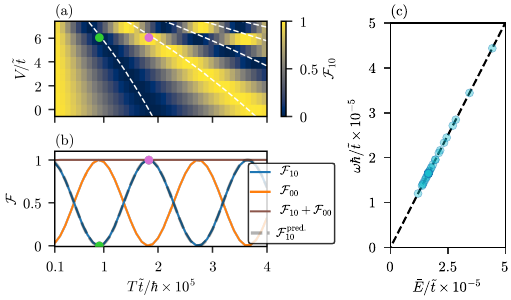}
    \caption{(a) X-gate braid fidelity, $\mathcal{F}_{10}=|\langle 1|0(T)\rangle|^2$ as a function of the strength of an impurity inserted at $\vec{r}=(4,0)$, and the braid time $T$. The white dashed lines show the predicted maxima and minima of the fidelity. (b) Line-cut of (a) at $V/\tilde{t}=5.8$. The information is shown to remain within the low-energy subspace but the fidelity oscillates as a function of braid time with the frequency $\add{\omega=\bar{E}/\hbar}$. Purple and green dots highlight common points in (a) and (b). Brown line shows total probability $\mathcal{F}_{00}+\mathcal{F}_{10}$.
    (c) Frequency $\omega$ of measured oscillations vs.\ $\bar{E}$ for impurities of various strengths and positions, which coincide with the predicted frequency $\add{\omega=\bar{E}/\hbar}$.}
    \label{fig:oscillations}
\end{figure}

In Fig.\,\ref{fig:oscillations}\,(a), we show the results of many-body simulations for an X-gate with an impurity at $\vec{r}=(4,0)$.
For $V=0$, the fidelity oscillates with frequency $\add{\omega}=\bar{E}/\hbar$, following the formula in\,\cite{hodge2024}, $\mathcal{F}_{10}=\cos^2(\bar{E}T/2\hbar)$.
As the impurity strength increases, the oscillation frequency also increases.
The predicted maxima and minima, calculated using the average energy, are shown as dashed white lines. As $\bar{E}$ depends non-trivially on $V$, these curves are found by identifying appropriate braid times, $T$, which satisfy $\bar{E}T=n\pi$ for a given impurity strength, where $n\in\mathbb{Z}^+$.

The oscillations for $V=5.8\tilde{t}$ are shown in Fig.\,\ref{fig:oscillations}\,(b).
 Fig.\,\ref{fig:oscillations}\,(c) plots the observed frequencies of the oscillations in the fidelity for different sites and impurity strengths against the $\bar{E}$ of each of their braids. Although $\bar{E}$ is site dependent, all points lie along the line $\add{\omega=\bar{E}/\hbar}$. 
This confirms $\bar{E}$ as a good measure for the quality of braids. We conclude that a single impurity effects the average energy $\bar{E}$ of a braid, which then leads to braiding errors as predicted in Ref.\,\onlinecite{hodge2024}. In other words, a single impurity affects the braiding of MZMs only through influencing the dynamical hybridization between MZMs.

\section{Random disorder}

\begin{figure}[ht]
    \centering
    \includegraphics[width=\columnwidth]{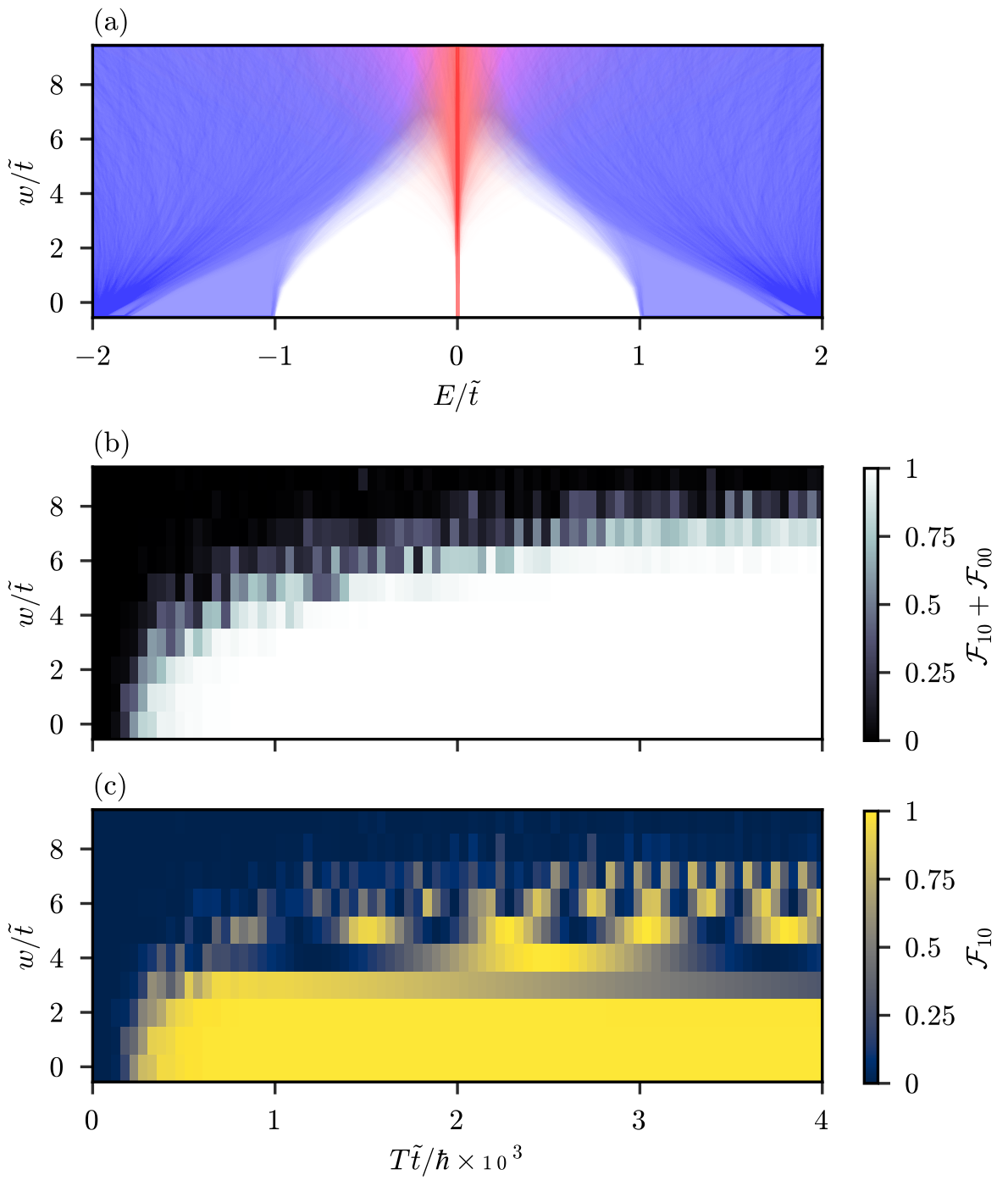}
    \caption{X-gate braid with random disorder. (a) The energy range $E_i^{\rm max}(t)-E_i^{\rm min}(t)$ of the $i$th energy level during the braid is plotted as a shaded region for increasing disorder strength $w$. The results of 100 braids are overlayed so that the opacity corresponds to the likelihood of a state being within that range at some point during the braid. The MZMs 
    in red, all bulk states in blue. Parameters used: $\mu_{\rm topo}=0.05\tilde{t}$, $\mu_{\rm triv}=10\tilde{t}$. Other values of 
    $\mu_{\rm topo}$ are shown in App.\,\ref{app:disorder}.
     (b) Occupation of the low-energy subspace $\mathcal{F}_{00}+\mathcal{F}_{10}$
     at the end of the braid for a single impurity distribution. As the impurity range increases, the minimum adiabatic braid time increases. (c) Corresponding fidelity $\mathcal{F}_{10}$ for the same impurities as in (b). The fidelity is constrained by the groundstate occupancy, and the oscillation frequency increases as the disorder strength increases showing the interplay between the two timescales.
    }
    \label{fig:disorder-gap} 
\end{figure}

The topological phase of superconducting nanowire or chain systems are robust to adding a disorder potential\,\cite{brouwerPRB2011,hegde2016,degottardiPRL2013,degottardiPRB2013,gergs2016,crawford2020,lieu2018}.
Common wisdom tells us that when the disorder strength reaches a critical value, the topological gap closes and the system becomes topologically trivial. Indeed,
this can be seen, for example, in Fig\,\ref{fig:disorder-gap}\,(a), where the gap for 100 disorder configurations closes around a disorder strength of $w=7\tilde{t}$.
When the gap becomes smaller, it takes a longer braid time to get to the adiabatic limit for typical braids.
Fig.\,\ref{fig:disorder-gap}\,(b) shows the total probability of the ground state to be in the low-energy subspace after a braid for a representative disorder configuration.
As the disorder strength increases, the region for the state to have 100\% probability in the low-energy subspace takes longer braid times.
As the gap closes, the MZMs start to hybridize more since the coherence length increases\,\cite{cheng2009}.
This can be seen as an energy splitting in the red lines in Fig.\,\ref{fig:disorder-gap}\,(a).
This energy splitting causes the braiding result to oscillate, as seen in Fig.\,\ref{fig:disorder-gap}\,(c).


In Refs.\,\onlinecite{gergs2016,lieu2018} 
it was pointed out that disorder can also have a contrary effect: for the Kitaev chain or extensions thereof as a prototype of a one-dimensional topological superconductor, disorder was shown to enhance the topological phase and localize the MZMs more strongly. We will first revisit these previous results and then show how it affects the braiding dynamics and quality of braids.

\begin{figure}[t!]
    \centering
    \includegraphics[width=\columnwidth]{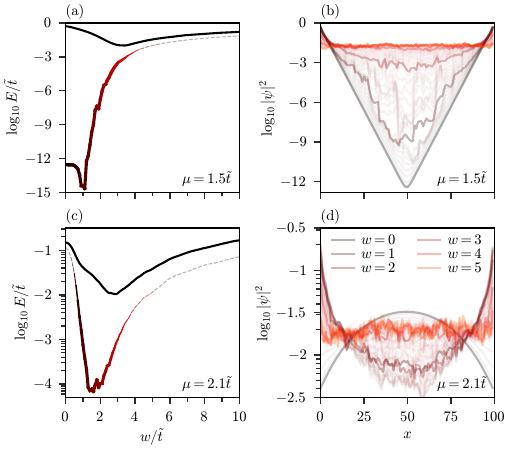}
    \caption{Disorder enhanced topological phases. The MZM energies $E$ (a, c) and wavefunctions $|\psi|^2$ (b, d) of the Kitaev chain (100 sites, $\Delta=\tilde{t}$) as a function of disorder $w$. Averaged over 1000 disorder realizations.
    Parameters used: (a, b) $\mu=1.5\tilde{t}$ and (c, d) $\mu=2.1\tilde{t}$. Disorder strength is indicated by color, while the thickness of the lines in (a) and (c) show how likely the chain is to be topological (thick=topological, thin=trivial), determined by the number of transfer matrix eigenvalues with magnitude less than 1.}
    \label{fig:disorder+topology}
\end{figure}

For Kitaev chains with near-zero chemical potential the energy quickly increases from essentially zero to finite values for disorder greater than $w=\tilde{t}$, however there is a reduction in energy for chemical potentials closer to $\mu_c$. In Fig.\,\ref{fig:disorder+topology} we show the static energy of MZMs for the Kitaev chain along with the energy of the bulk gap as a function of disorder strength $w$, averaged over many disorder realizations. In Fig.\,\ref{fig:disorder+topology}\,(a) there is a reduction in energy splitting compared to the clean chain for small disorder strength, up to a minimum $w_{\rm min}=1.1\tilde{t}$, after which the energy rapidly increases. Using the transfer matrix approach to describe low energy states \cite{degottardiPRL2013} we can determine whether a specific disordered chain is topological or trivial by counting the number of eigenvalues with magnitude less than 1 as outlined in Ref.\,\cite{gergs2016}. The likeliness that a chain hosts MZMs for a given disorder strength is indicated by the thickness of the curve in Fig.\,\ref{fig:disorder+topology} (a,c). A dashed guide is provided for the trivial chains when the line becomes too thin to see. In Fig.\,\ref{fig:disorder+topology}\,(b) the wavefunctions $|\psi|^2$ of the MZMs (plotted on a log scale) are shown along the chain position $x$. Exponentially localized wavefunctions can clearly be distinguished from less localized ones. For the chosen parameters which yield nearly perfectly localized MZMs in the absence or presence of low disorder, disorder does not have any positive effect on the energy of MZMs and the localization of their wavefunctions. However, in Fig.\,\ref{fig:disorder+topology}\,(c) we have set $\mu=2.1\tilde{t}$ so that without disorder the chain is in a trivial state which should be incapable of hosting MZMs. Increasing disorder \add{pushes the system back into the topological phase and} leads to a drastic reduction of the MZM-energy which reaches a minimum at $w_{\rm min}=1.6\tilde{t}$ and then increases again. Please note that $w_{\rm min}$ is not universal and depends on the other parameters. In Fig.\,\ref{fig:disorder+topology}\,(d) we show again the wavefunctions $|\psi|^2$ confirming how disorder forces the wavefunctions to be more localized as compared to the disorder-free case. Thus we confirm the earlier findings of Refs.\,\onlinecite{gergs2016,lieu2018}.

\begin{figure}
    \centering
    \includegraphics[width=\columnwidth]{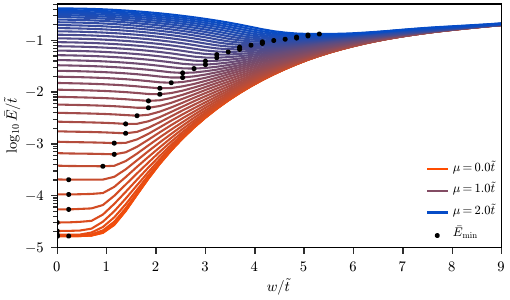}
    \caption{Average energy $\bar{E}$ of an X-gate braid vs.\ disorder $w$.
    Averaged over 100 disorder realizations. Each line corresponds to a different value of $\mu_{\rm topo}$, ranging from $\mu_{\rm topo}=0$ in blue to $\mu = 2\tilde{t}$ (critical point at $w=0$) in red. The minimal average energy, $\bar{E}_{\rm min}$, for each curve is identified with a black marker.}
    \label{fig:Ebar-vs-disorder}
\end{figure}

After these static considerations, now we investigate how the braiding dynamics and braiding errors of a Majorana qubit on a T-junction will be effected by disorder. In Fig.\,\ref{fig:Ebar-vs-disorder}, the average energy $\bar{E}$ vs.\ disorder $w$ is shown for various values of the topological chemical potential from $\mu_{\rm topo}=0$ (perfectly localized case) to $\mu=2\tilde{t}$ (critical point between topological and trivial phase) in increments of $0.05\tilde{t}$. For almost-perfectly localized MZMs, i.e., small values of $\mu_{\rm topo}$, the effect of moderate disorder is negligible.
\add{
The predominant energy splitting occurs only through the trivial region; 
thus any hybridization within the topological segments is obfuscated. Since the trivial segments dominate the hybridization for values of $\mu_{\rm topo}$ close to zero, $\bar{E}$ does not decrease with $w$ (see Fig.\,\ref{fig:Ebar-vs-disorder}).
For larger values of disorder $w$, $\bar{E}$ drastically increases and damages the braiding performance.}
\add{However, for slightly larger $\mu_{\rm topo}$ values, we observe that the average energy is minimized at a non-zero disorder strength (indicated by black dots in Fig.\,\ref{fig:Ebar-vs-disorder}).}
For instance, for $\mu=\tilde{t}$ we find the minimal average energy $\bar{E}_{\rm min}$ for $w=2.3\tilde{t}$. 
\add{
To understand this reduction of $\bar{E}$ for finite $w$, it is crucial to understand that both the trivial and topological segments of the T-junction contribute to the average energy $\bar{E}$. However, disorder mostly affects the topological region. Only for parameters where the hybridization of the topological segments dominates, disorder can be beneficial and reduce this hybridization, and subsequently $\bar{E}$.
Thus only for larger values of $\mu_{\rm topo}$, is it possible to observe a measurable reduction in $\bar{E}$, and the minimum $\bar{E}_{\rm min}$ moves to larger values of $w$.}
 \add{As $\mu_{\rm topo}$ approaches a substantial proportion of $\mu_c$,} this minimum energy can be reached for quite sizable disorder strengths, such as $w=4\tilde{t}$ or $5\tilde{t}$; however, such parameters cannot be used for braiding because the bulk gap is already too small or even closed by the disorder (Fig.\,\ref{fig:disorder-gap}(a)). We see that the disorder-enhanced topological phase is limited to some range of disorder strength, \add{as we explicate further below.}
To summarize this paragraph, as for the static case, also the dynamical analysis suggests that there is a window of opportunity where disorder might indeed be able to reduce braiding errors and improve braiding outcomes.

\begin{figure}[t!]
    \centering
    \includegraphics[width=\columnwidth]{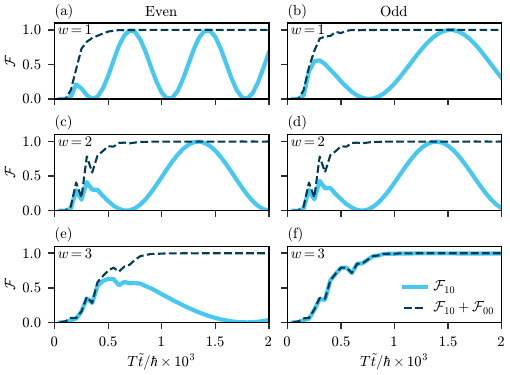}
    \caption{Fidelity $\mathcal{F}_{10}$ and ground state occupation of MZMs, $\mathcal{F}_{00}+\mathcal{F}_{10}$, as a function of total braid time $T$.
    X-gate braid outcomes for a single disorder configuration at varying strengths. The solid curves show the fidelity of the braid, while the dashed lines show the total groundstate occupation. (a, b) have disorder strength $w=\tilde{t}$, (c, d) have disorder strength $w=2\tilde{t}$, and (e, f) have disorder strength $w=3\tilde{t}$. (a, c, e) correspond to the even parity while (b, d, f) to odd parity.}
    \label{fig:diabatic-effects}
\end{figure}

So far we have only focused on the effect of disorder on $\bar{E}$ and the fidelity oscillations caused by too large $\bar{E}$\,\cite{hodge2024}. With other words, we have only considered the upper bound of the speed limit of Majorana braiding. In the following, we also need to consider the lower limit which is due to diabatic effects. That is, by braiding too fast transitions from the Majorana ground state manifold to excited states are possible, spoiling successful braiding outcomes. A shrinking bulk gap obviously amplifies such diabatic effects. In Fig.\,\ref{fig:diabatic-effects}, we have tried to quantify how diabatic effects mess with the previously identified window of opportunity of disorder-enhanced braiding outcomes.

In the following, we consider both the even and odd parity subspaces of a Majorana qubit (consisting of four MZMs). We perform an X-gate on the initialized state $\ket{0}_L$. A braid is thus characterized by the fidelity $\mathcal{F}_{01}=|\braket{0 | 1(T)}|^2$ as shown in Fig.\,\ref{fig:diabatic-effects}, with $\mathcal{F}_{01}=1$ indicating success. In addition, we also show the total probability to stay in the MZM ground state sector, $|\braket{0 | 0(T)}|^2 + |\braket{0 | 1(T)}|^2$, as the dashed black line in Fig.\,\ref{fig:diabatic-effects}. A total probability smaller than 1 indicates the diabatic regime, where transitions into excited (bulk) states will spoil any controlled braiding attempts.

Interestingly, however, when it comes to the stability of the braiding outcome, the odd parity subspace has been shown to reveal a potential advantage over the even subspace\,\cite{bedow-23,hodge2024}.
This is shown in the hybridizing regime, where the angular frequency of oscillation is given by $\hbar\omega = \bar{E}_{12}+\bar{E}_{13}+(-1)^p(\bar{E}_{34}+\bar{E}_{24})$, where $\bar{E}_{ij}$ is the average hybridization between $\gamma_i$ and $\gamma_j$ and $p=0\ (1)$ for even (odd) parity.
This relative difference is significant, as it may allow some splittings to cancel out, leading to greater periods of oscillation.
This can be seen in Fig.\,\ref{fig:diabatic-effects}. An ideal situation would be that the maximum of the fidelity oscillations (blue solid curve) is broad corresponding to a large window of successful braiding times, and at the same time the total probability (black dashed curve) reaches one rather quickly. Where the total probability takes too long to reach 1, the first maximum of the fidelity oscillations will be spoiled, as can be seen for Fig.\,\ref{fig:diabatic-effects}(a--e). Only for considerable disorder strength $w=3\tilde{t}$ we obtain for the odd parity subspace a scenario where the first fidelity maximum is so broad that the diabatic effects do not influence the braiding performance anymore: in this particular example, any total braid time between $1000\hbar/t$ and $2000\hbar/t$ will lead to a successful braid. The same system but with less disorder would, however, not result in a successful braid. While not ubiquitous or universal, within the regime of disorder-enabled topological phases are also regimes where we find disorder-enabled braiding success.


%
%
\section{Conclusion}

We find that the non-Abelian exchange properties of MZMs are robust in the presence of disorder.
This was shown in a T-junction of Kitaev chains for single impurities, as well as random disorder. In the case of a single impurity, the hybridization between MZMs increases, leading to faster oscillations around the Bloch sphere, i.e., an increase in braiding errors, as described in Ref.\,\onlinecite{hodge2024}.
\add{For strong random disorder, the hybridization experiences a similar increase, and the introduction of many low energy YSR states causes a reduction in the effective gap.}
Not only does this cause oscillations, but also increases the time scales necessary for adiabatic braiding.
As pointed out previously for the static case, disorder can also have a positive effect and stabilize or even increase the topological phase. We show that this disorder-enhanced topological phase due to disorder extends from the static to the dynamical case and can indeed lead to a better braiding performance of MZMs. As shown by our analysis, the increasing disorder leads to a monotonous shrinking of the bulk gap, thus amplifying diabatic effects, which is antagonistic to any braiding success. Nevertheless, we show that disorder is not always destructive, but that there remains a window of opportunity where disorder can help to improve braiding outcomes.

These simple investigations of the Kitaev chain or T-junction give an interesting view regarding the role of disorder.
However, a more thorough investigation must clarify how universal these results are. In particular, if they carry over to spinful and multi-orbital models. Moreover, one should consider the type of disorder relevant to a specific platform.
This could be charge or magnetic impurities, correlated disorder, disorder in the hopping, disorder in the proximity induced gap, changes in effective mass, $1/f$ noise, just to mention the most obvious candidates.
Another interesting aspect would be to consider self-consistency in the superconducting order parameter and the electric potential, both of which could have significant impact on the robustness of braiding. 
Our results show that simple models of disorder can have interesting and unexpected results, and lays the foundation for future studies of braiding with disorder.  


\begin{acknowledgments}
S.R.\ acknowledges support from the Australian Research Council through Grant No.\ DP200101118 and DP240100168.
This research was undertaken using resources from the National Computational Infrastructure (NCI Australia), an NCRIS enabled capability supported by the Australian Government.
\end{acknowledgments}

\appendix

\section{\label{app:subgap}Subgap states}

The energy level of the YSR state introduced by a single impurity, $E_b$, can be determined through the use of a T-matrix calculation. 
The T-matrix, $T(\vec{p}_1,\vec{p}_2,\omega)$, is defined such that we may decompose Green's function, $G(\vec{p}_1,\vec{p}_2,\omega)$, as $G(\vec{p}_1,\vec{p}_2,\omega) = G_0(\vec{p}_1,\omega) T(\vec{p}_1,\vec{p}_2,\omega) G_0(\vec{p}_2,\omega)$ where $G_0(\vec{p},\omega)$ is the Green's function of the clean system in terms of the momentum, $\vec{p}$ and frequencies, $\omega$. 
Here, the T-matrix contains the information of all processes that involve scattering off the impurity. 
It is therefore possible to identify bound states that constantly interact with the impurity with singularities in the T-matrix that coincide spatially with the impurity \cite{sau2013a,kaladzhyan2016,kaladzhyan2019}.
For an infinite Kitaev chain with a single delta-like charge impurity, exact calculations for $\mu=0$ were reported\,\cite{kaladzhyan2019}:
\begin{equation}\label{Eb-old}
\begin{split}
   &E_b=\\
   &\pm\!\!\sqrt{8\left(\tilde{t}^2+\Delta^2+\frac{V^2}{4}\pm\sqrt{\left(\tilde{t}^2+\Delta^2+\frac{V^2}{4}\right)^2-4\tilde{t}^2\Delta^2}\right)}
   \end{split}
\end{equation}
\add{Using the results for finite $\mu$ from Ref.\,\cite{kaladzhyan2019}, we were able to obtain the following expression (limited to $|\Delta|=\tilde{t})$}: 
\begin{equation}\label{Eb-new}
 E_b = \pm\frac{\alpha \pm \sqrt{\beta}}{\mu-V}
\end{equation}
with $\alpha$ and $\beta$ defined below:
\begin{gather}
    \alpha = \frac{1}{2}V^2-\mu V+\mu^2\ , \\
    \beta = \frac{1}{4}V^4-\mu V^3+(\mu^2+4\tilde{t}^{\,2})V^2-8\mu \tilde{t}^{\,2} V+ 4\mu^2\tilde{t}^{\,2}\ .
\end{gather}
Both\eqref{Eb-old} and \eqref{Eb-new} have four solutions, two of which descend from the bulk and form the subgap states while the other two raise in energy further away from the MZMs and thus we are not concerned with them. Furthermore, the subgap states only exist for $V\in[0,2\mu]$ for the trivial regions where $|\mu|>2\tilde{t}$; for topological chains, i.e., that $|\mu|<2\tilde{t}$, the subgap states exist only for $V\in(-\infty,0]\cup[2\mu,\infty)$.

\begin{figure}[b!]
    \centering
    \includegraphics[width=\columnwidth]{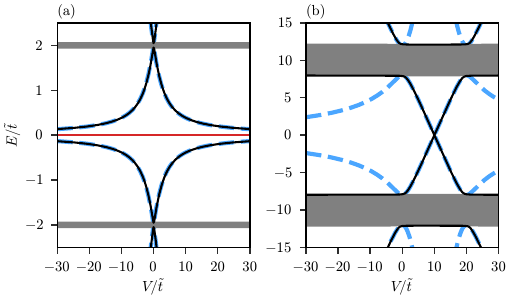}
    \caption{Energies due to an impurity on an infinite chain (blue dashed lines) vs.\ finite T-junction (black solid line) as a function of impurity strength $V$. Bulk bands shown in dark grey. (a) Topological regime ($\mu=0.05\tilde{t}$), i.e., the entire T-junction is topological. MZMs are shown in red. (b) Trivial regime ($\mu=10\tilde{t}$).}
    \label{fig:boundstates}
\end{figure}

In Fig.\,\ref{fig:boundstates} these analytical solutions for the infinite chain are compared to the energy spectra obtained by diagonalizing the Hamiltonian of the T-junction at t=0 with a single impurity of strength $V$. Fig.\,\ref{fig:boundstates}\,(a) places the impurity at $\vec{r}=(-10,0)$ in the topological region while Fig.\,\ref{fig:boundstates}\,(b) has the impurity placed at $\vec{r}=(-4,0)$. 

As mentioned in Sec.\,\ref{sec: single impurities}, strong impurities in the topological region generate subgap states with energies that are inversely proportional to the impurity strength,  $E_b\approx\pm4\tilde{t}|\Delta|/V$ \cite{kaladzhyan2019}.

In both cases the analytic results for the infinite chain (blue dashes) strongly agree with the numeric results for the finite T-junction (black) within the regions of validity. 
This, however, does not hold when the impurities are placed directly next to boundaries. 
In Fig.\,\ref{fig:braid}\,(b) as the subgap state doesn't appear until step 2, even though the impurity is within the topological region for the whole braid. 
\add{This is due to its location on the border between different regions, so when the impurity strength is sufficient such that the site becomes trivial, instead of forming a defect within the topological region it instead increases the length of the trivial state, avoiding the introduction of a subgap state. 
This effect also occurs at the ends of the legs where the absence of a site can in some respects be considered as the ideal trivial region.}

Aside from the subgap state caused by the phase difference between horizontal and vertical legs, the braiding protocol introduces additional subgap states. This is due to the boundary between regions being broadened from an instantaneous transition from $\mu_{\rm topo}$ to $\mu_{\rm triv}$ across adjacent sites. This deformation to the chemical potential can be considered to be a spatially extended impurity. This also generates a bound state, however as seen in Fig.\,\ref{fig:braid}\,(b) these barely descend into the bulk and thus are not expected to significantly impact the braid results.

\section{\label{app:disorder}Disorder-dependence of energies of an X-gate braid}

\begin{figure}[t!]
    \centering
    \includegraphics[width=\columnwidth]{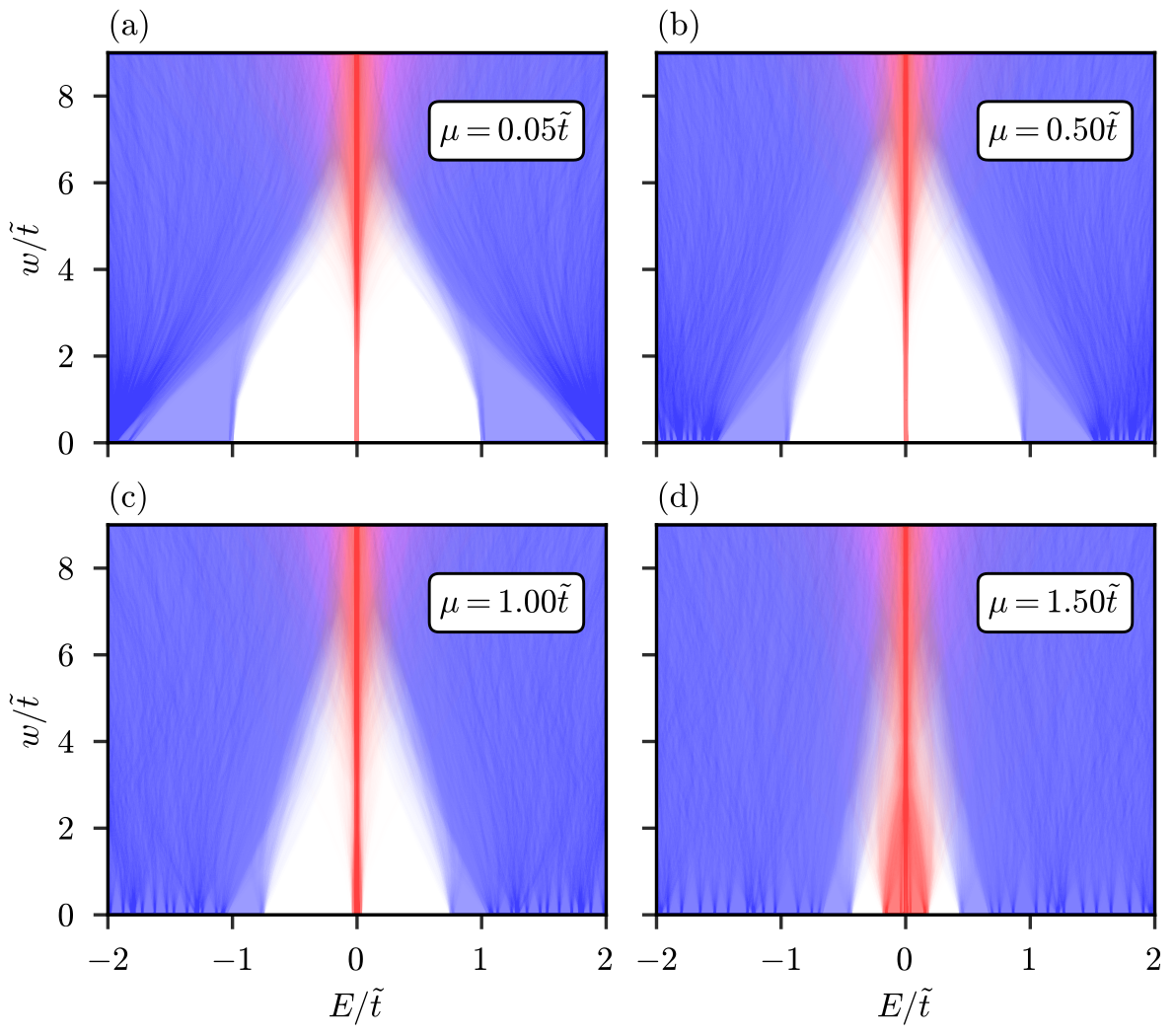}
    \caption{Same as Fig.\,4\,(a) but for parameters (a) $\mu_{\rm topo}=0.05$, (b) $\mu_{\rm topo}=0.5$, (c) $\mu_{\rm topo}=1.0$ and (d) $\mu_{\rm topo}=1.5$. The MZMs are shown in red, and the bulk states in blue. In all panels $\mu_{\rm triv} = 10\tilde{t}$.}
    \label{fig:more-disorder}
\end{figure}

In Fig.\,\ref{fig:more-disorder} we show more disorder results for an X-gate braid in the presence of random disorder. In particular, the energy range $E^{\rm max}_i(t)-E^{\rm min}_i(t)$ of the $i$th energy level during
the braid is plotted as a shaded region for increasing disorder
strength $w$. In each panel, the results of 100 braids are overlayed so that the
opacity corresponds to the likelihood of a state being within
that range at some point during the braid. Please note that Fig.\,\ref{fig:more-disorder}\,(a) ($\mu_{\rm topo}=0.05$) is identical to Fig.\,\ref{fig:disorder-gap}\,(a).

\bibliography{references}

\end{document}